\documentclass[onecolumn,superscriptaddress, notitlepage,amssymb, nobibnotes,nofootinbib,longbibliography]{revtex4-1}

\usepackage[letterpaper,top=2cm,bottom=2cm,left=3cm,right=3cm,marginparwidth=1.75cm]{geometry}
\usepackage{float}
\usepackage{amsmath}
\usepackage{amssymb}
\usepackage{ulem}
\usepackage{graphicx}
\usepackage[colorlinks=true, allcolors=blue]{hyperref}
\usepackage[dvipsnames]{xcolor}
\usepackage{makecell}
\usepackage{longtable}
\usepackage{rotating}

\newcommand{\scat}{\ensuremath{\mathcal{S\!\ }}}
\newcommand{\absb}{\ensuremath{\mathcal{A\!\ }}}
\newcommand{\tran}{\ensuremath{\mathcal{T\!\ }}}
\newcommand{\F}{\ensuremath{\mathcal{F\!\ }}}
\newcommand{\ROC}{\ensuremath{R\ }}

\makeatletter
\def\@ssect@ltx#1#2#3#4#5#6[#7]#8{%
  \def\H@svsec{\phantomsection}%
  \@tempskipa #5\relax
  \@ifdim{\@tempskipa>\z@}{%
    \begingroup
      \interlinepenalty \@M
      #6{%
       \@ifundefined{@hangfroms@#1}{\@hang@froms}{\csname @hangfroms@#1\endcsname}%
       {\hskip#3\relax\H@svsec}{#8}%
      }%
      \@@par
    \endgroup
    \@ifundefined{#1smark}{\@gobble}{\csname #1smark\endcsname}{#7}%
  }{%
    \def\@svsechd{%
      #6{%
       \@ifundefined{@runin@tos@#1}{\@runin@tos}{\csname @runin@tos@#1\endcsname}%
       {\hskip#3\relax\H@svsec}{#8}%
      }%
      \@ifundefined{#1smark}{\@gobble}{\csname #1smark\endcsname}{#7}%
      \addcontentsline{toc}{#1}{\protect\numberline{}#8}%
    }%
  }%
  \@xsect{#5}%
}%
\makeatother

\begin{document}
\title{Micro-fabricated mirrors with finesse exceeding \\one million}

\author{Naijun~Jin} 
\affiliation{Department of Applied Physics, Yale University, New Haven, CT 06520, USA}
\author{Charles~A.~McLemore}
\affiliation{Department of Physics, University of Colorado Boulder, Boulder, CO 80309, USA}
\affiliation{National Institute of Standards and Technology, Boulder, CO 80305, USA}
\author{David~Mason}
\affiliation{Department of Applied Physics, Yale University, New Haven, CT 06520, USA}
\author{James~P.~Hendrie}
\affiliation{Department of Physics, University of Colorado Boulder, Boulder, CO 80309, USA}
\affiliation{National Institute of Standards and Technology, Boulder, CO 80305, USA}
\author{Yizhi~Luo}
\affiliation{Department of Applied Physics, Yale University, New Haven, CT 06520, USA}
\author{Megan~L.~Kelleher}
\affiliation{Department of Physics, University of Colorado Boulder, Boulder, CO 80309, USA}
\affiliation{National Institute of Standards and Technology, Boulder, CO 80305, USA}
\author{Prashanta~Kharel}
\affiliation{Department of Applied Physics, Yale University, New Haven, CT 06520, USA}
\author{Franklyn~Quinlan}
\affiliation{National Institute of Standards and Technology, Boulder, CO 80305, USA}
\author{Scott~A.~Diddams}
\affiliation{Department of Physics, University of Colorado Boulder, Boulder, CO 80309, USA}
\affiliation{Department of Electrical, Computer and Energy Engineering, University of Colorado Boulder, Boulder, CO 80309, USA}
\affiliation{National Institute of Standards and Technology, Boulder, CO 80305, USA}
\author{Peter~T.~Rakich}
\affiliation{Department of Applied Physics, Yale University, New Haven, CT 06520, USA}
\date{March 29, 2022}
\begin{abstract}
The Fabry-Pérot resonator is one of the most widely used optical devices, enabling scientific and technological breakthroughs in diverse fields including cavity QED, optical clocks, precision length metrology and spectroscopy. 
Though resonator designs vary widely, all high-end applications benefit from mirrors with the lowest loss and highest finesse possible. Fabrication of the highest finesse mirrors relies on centuries-old mechanical polishing techniques, which offer losses at the part-per-million (ppm) level.
However, no existing fabrication techniques are able to produce high finesse resonators with the large range of mirror geometries needed for scalable quantum devices and next-generation compact atomic clocks. 
In this paper, we introduce a new and scalable approach to fabricate mirrors with ultrahigh finesse ($\geq10^6$) and user-defined radius of curvature spanning four orders of magnitude ($10^{-4}-10^{0}$ m).
We employ photoresist reflow and reactive ion etching to shape and transfer mirror templates onto a substrate while maintaining sub-Angstrom roughness.
This substrate is coated with a dielectric stack and used to create arrays of compact Fabry-Pérot resonators with finesse values as high as 1.3 million and measured excess loss ~$<$ 1 ppm.
Optical ringdown measurements of 43 devices across 5 substrates reveal that the fabricated cavity mirrors—with both small and large radii of curvature—produce an average coating-limited  finesse of 1.05 million. 
This versatile new approach opens the door to scalable fabrication of high-finesse miniaturized Fabry-Pérot cavities needed for emerging quantum optics and frequency metrology technologies.
\end{abstract}

\maketitle

\section*{Main}

Among optical resonators, high-finesse Fabry-Pérot cavities produce unrivaled frequency stability, quality factors, and power handling, enabling scientific and technological breakthroughs in a broad range of applications\cite{thompson_observation_1992,reiserer_cavity-based_2015,mckeever_deterministic_2004,colombe_strong_2007, matei_15um_2017, fortier_generation_2011}.
For the next generation in quantum communications, computation, and time-keeping systems, it will be necessary to bring these performance advantages to compact, integrated platforms\cite{newman_architecture_2019,maurice_miniaturized_2020,spencer_optical-frequency_2018}.  This will require a scalable fabrication technique that is flexible enough to meet the varied demands of disparate applications.
Many applications benefit from increased finesse, which translates to larger intracavity fields, increased storage times, and narrower linewidths.
But geometry can be equally important, as the optimal mode volumes and spot size can vary dramatically for different applications, placing different requirements on the mirror radius of curvature (\ROC$\!\!$).
In quantum optics, where the cooperativity between single atoms and optical resonators scales inversely with mode area\cite{hunger_fiber_2010}, microcavity geometries with small radius of curvature ($R \sim 10^{-4}-10^{-2}$ m) are desirable.
Conversely, for ultra-stable reference cavities in timekeeping applications, frequency noise can be minimized by averaging over thermal fluctuations with large mode areas\cite{levin_internal_1998}, requiring a large radius of curvature ($R\sim 1$ m).

To maximize finesse (\F), it is necessary to minimize all sources of optical loss within the cavity.
This is seen from the definition, ${\F=\pi/(\tran+\absb+\scat)}$, where \tran, \absb, and \scat represent the fractional energy loss (per mirror) resulting from transmission, absorption, and scattering, respectively.
Thus, an ultrahigh-finesse resonator ($\F>10^6$) requires $\tran+\absb+\scat$ to be at the few ppm level.
Using ion-beam sputtering deposition techniques, highly uniform dielectric coatings with absorptive losses (\absb) of $\sim$~1 ppm are available\cite{rempe_measurement_1992}.
However, roughness on the mirror surface and subtle imperfections in the mirror shape can both contribute to unwanted scattering losses, resulting in stringent requirements on the surface quality of the mirror template.  
For example, at telecom wavelengths, a mirror template with an ideal surface profile (i.e., without any low spatial frequency shape imperfections) must have sub-Angstrom RMS surface roughness to achieve scattering losses (\scat) below 1 ppm.

\begin{figure*}[t]
\begin{center}
\includegraphics[width=1.0\textwidth]{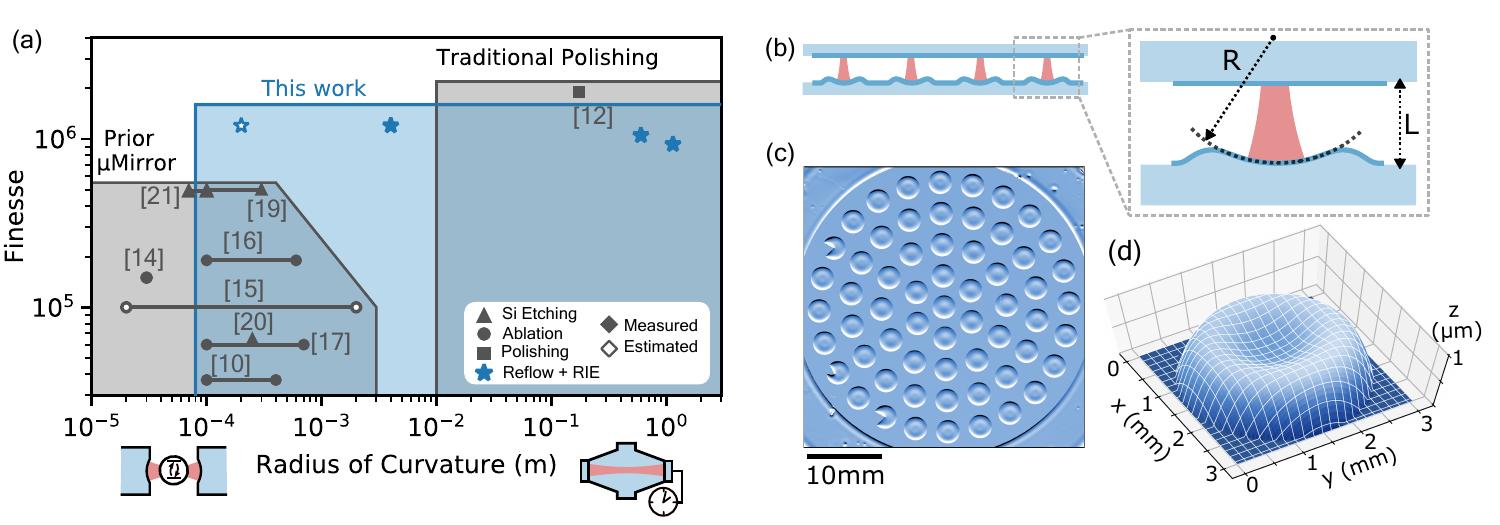}
\caption{{\bf Micromirror Solutions}
{\bf (a)} Survey of micromirror fabrication techniques.  The shaded regions and points illustrate the achievable geometry (radius of curvature) and finesse of different techniques. Grey corresponds to prior work including laser ablation (circles), isotropic chemical etching (triangles) and traditional polishing. Blue corresponds to this work. 
Filled points indicate measured finesses, while unfilled points indicate fabricated mirror templates with finesse predicted based on surface scattering estimates/simulations.
Note that these reference finesse values were measured at different optical wavelengths, which will modify the impact of surface scattering. Further details on the literature values are available in Supplementary Section S1.
{\bf (b)} Illustration of the cavities built in this work.
{\bf (c)} Image of an array of 58 reflowed photoresist disks on a two-inch wafer that can be etched to make mirror templates. Non-circular devices on the outer ring are intended for alignment purposes.
{\bf (d)} Measured profile of a fabricated mirror template with $R\approx 1$ m.}

\label{f:F1}
\end{center}
\end{figure*}

Specialized chemical-mechanical polishing techniques, sometimes referred to as super-polishing\cite{nelson_creating_2019}, are used to meet these stringent requirements on individually polished discrete mirror components.
This polishing technique can achieve the necessary sub-Angstrom roughness, but only for large radius of curvature mirrors (\ROC$\sim$ 10 mm$-$1000 mm). 
Motivated by quantum optics, new fabrication techniques utilizing laser ablation of glass\cite{hunger_fiber_2010,muller_ultrahigh-finesse_2010,hunger_laser_2012,uphoff_frequency_2015,takahashi_novel_2014}, and chemical etching of silicon\cite{trupke_microfabricated_2005,wachter_silicon_2019,biedermann_ultrasmooth_2010, fait_high_2021} have been developed in recent years, finding applications in a wide range of experiments\cite{colombe_strong_2007,merkel_coherent_2020,takahashi_strong_2020,steiner_single_2013,kashkanova_superfluid_2017,flowers-jacobs_fiber-cavity-based_2012,janitz_fabry-perot_2015,albrecht_coupling_2013}. 
While these new techniques have the potential for scalable fabrication, they are limited to the production of small $R$ ($\lesssim 1$ mm) mirrors\cite{ott_millimeter_long_2016, takahashi_novel_2014}, with finesse values that fall short of traditional polishing techniques (see comparison in Fig 1(a)).  
Thus, it remains an outstanding challenge to identify a scalable fabrication technique that yields ultrahigh-finesse mirrors, with access to both small and large mode volumes.

In this paper, we demonstrate a wafer-scale fabrication technique that produces ultrahigh-finesse ($\geq10^6$) mirrors with a user-defined $R$ spanning from 100 microns to 1 meter, necessary to satisfy the demanding needs of applications ranging from quantum optics to low-noise laser oscillators.
Arrays of microfabricated mirrors are formed on a single substrate using a solvent-vapor based resist reflow process.  
Through this process, photoresist defines mirror shapes that are transferred into a substrate using an optimized dry etch, maintaining sub-Angstrom surface roughness.
Multilayer mirror coatings are then deposited, creating arrays of compact Fabry-Pérot resonators whose performance is evaluated using optical ring-down measurements.
Measurements of 43 devices across 5 substrates (with both small and large $R$) reveal that the fabricated cavity mirrors produce a mean (maximum) coating-limited finesse of 1.05 million (1.3 million), which, to the best of our knowledge, sets a record among micro-fabricated mirrors (and $R<10$mm mirrors in general).
This new  method thus enables the scalable production of compact Fabry-Pérot cavities with the state-of-the-art performance required by emerging technologies. 


\subsection*{Results}
\begin{figure*}[ht]
\begin{center}
\includegraphics[width=5in]{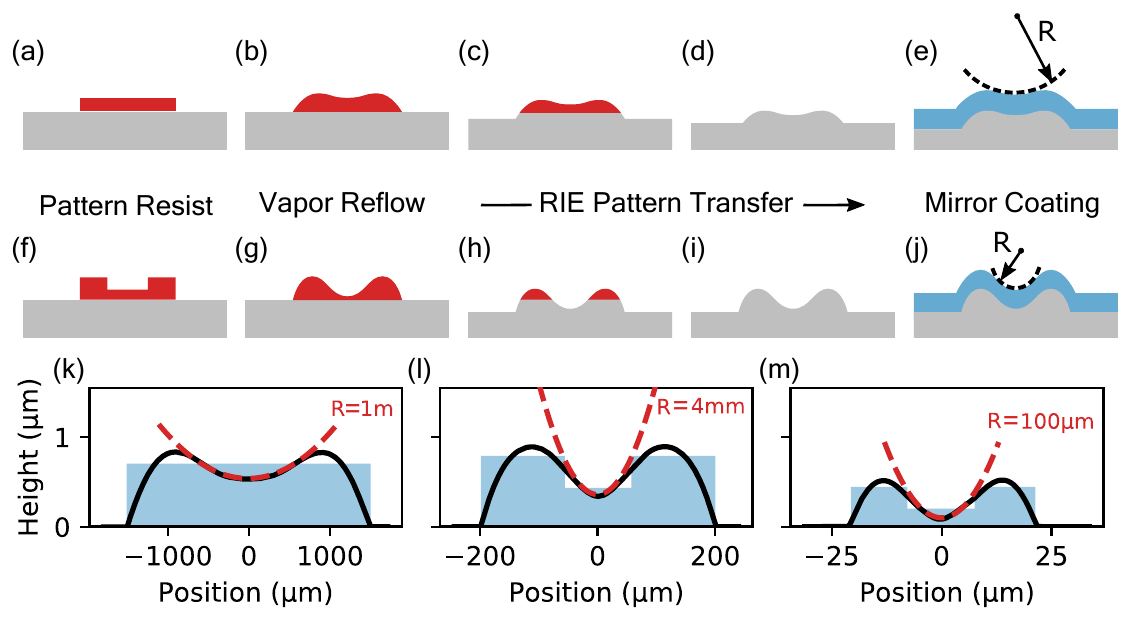}
\caption{{\bf Micromirror Fabrication}
The fabrication process begins with a single ({\bf a}) or multi-layer ({\bf f}) photoresist pattern.  After a timed solvent vapor reflow, a large ({\bf b}) or small({\bf g}) concave photoresist pattern is formed.  A reactive ion etch transfers this into the substrate ({\bf c,d,h,i}), before final application of mirror coating ({\bf e,j}).  Exemplary measured profiles (black) of reflowed structures with $R$ from 1 m to 100 $\mu$m are shown in {\bf (k-n)}. Illustrations of the approximate photoresist shape before reflow are shown in blue.
\label{f:F2}}
\end{center}
\end{figure*}

Through this fabrication approach, we use reflow techniques to create a resist profile that defines the shape of the mirror.
Photoresist patterns are first created on a super-polished substrate (e.g. fused silica) using UV lithography.
The single- and multi-level photoresist patterns, seen in Fig. 2a and Fig.~2f, are used to form large- and small-$R$ devices, respectively.
These resist patterns undergo reflow in a purpose-built solvent-vapor chamber; as the photoresist absorbs the solvent vapor, surface tension rounds any sharp corners as it seeks to minimize the surface area of the resist pattern\cite{emadi_vertically_2009}. 
In the limit of complete reflow, this disk is transformed into a dome\cite{kharel_ultra-high-q_2018, li_fabrication_2004}; however, for intermediate reflow times, a smooth parabolic surface is formed in the center of the resist pattern, as illustrated in Fig. 2b and Fig. 2g. An array of 58 such reflowed surfaces formed on a two-inch wafer is shown in Fig. 1c. 
When the photoresist pattern reaches the desired shape, the reflow process is halted, and the resist pattern is transferred into the substrate using an optimized reactive ion etch\cite{Li-smoothetch,minnick_optimum_2013}, as illustrated in Fig. 2c-d and 2h-i. 
Note that different etch rates for the photoresist and substrate result in a vertical rescaling of the pattern.
After these mirror templates are etched, a multilayer dielectric coating is deposited (Fig. 2e and 2j), producing an array of concave mirrors. 
Further details of the fabrication process can be found in Supplement Section S2.
Using this process, one can readily vary the mirror radius of curvature by 4 orders of magnitude through control of the initial photoresist geometry and reflow time.
Figure 2k-m show measured profiles of etched mirror templates (black) with radii of curvature ranging from $R=1$ m to $R=100$ $\mu$m; approximate resist profiles, used at the beginning of the fabrication process, are illustrated in blue.
Note that $R>1$ m and $R<100$ $\mu$m should be possible with modified photoresist patterns and techniques.
While the measured mirror curvature permits us to leverage Gaussian beam optics as the basis for resonator design, it is important to note that these mirror shapes deviate from a paraboloid at larger radial distances, and the mirrors have a finite size. 
Thus, in principle, the nontrivial surface profiles produced by the reflow process could contribute to clipping losses, limiting the performance of these mirrors. 
To investigate limitations posed by these shape-induced losses, we developed a numerical mode solver that builds on the techniques described in Refs\cite{Bouwmeester2010,Hunger2015}. 
Using a standard (e.g. Hermite-Gaussian) mode basis, this solver encodes a round-trip of optical propagation (including the exact mirror profile) into a mode scattering matrix.
This scattering matrix is then used to compute the eigenmodes of the resonator, including their associated loss rates. 
Simulating a plano-concave resonator geometry (Fig.~1b) using the measured mirror profile as the input, we find that the shape-induced diffractive losses of optimized mirror templates (Fig. 2f-j) are very small (i.e., $\mathcal{S}_{\mathrm{shape}}\leq 0.1$ ppm). 
In general, this low clipping loss is afforded by relatively deep mirror recesses, which produce large usable apertures. Further details on mirror depth and aperture constraints are available in Supplementary Section S2D.
Roughness induced scattering losses are perhaps the most significant barrier to realizing a finesse of greater than 1 million. One can show that the scattering loss associated with an RMS surface roughness, $\sigma_{\mathrm{rms}}$, is given by $\mathcal{S}_{\mathrm{rough}}=(4\pi\sigma/\lambda)^2$, where $\lambda$ is the wavelength of light\cite{bennett1961relation}. Hence, at $\lambda=1550$ nm, each mirror must have sub-Angstrom surface roughness ($\sigma\leq 1.2$~\AA) to meet the requirement $\mathcal{S}\leq 1$~ppm. Therefore, the etch process that transfers the photoresist pattern must not appreciably alter the roughness of the super polished substrate. For this task, we utilize a reactive ion etch that removes material primarily through ion bombardment (i.e., a physical etch) rather than chemical processes (see Supplementary Section S2E for further details). 
Figure \ref{f:F3}a shows a typical surface roughness measurement taken in the center of a microfabricated mirror; this measurement reveals an RMS surface roughness of 0.59\AA, corresponding to an estimated scattering loss of $\scat \cong 0.23$ ppm at 1550 nm wavelengths.  


\begin{figure*}[hhtt]
\begin{center}
\includegraphics[width=6in]{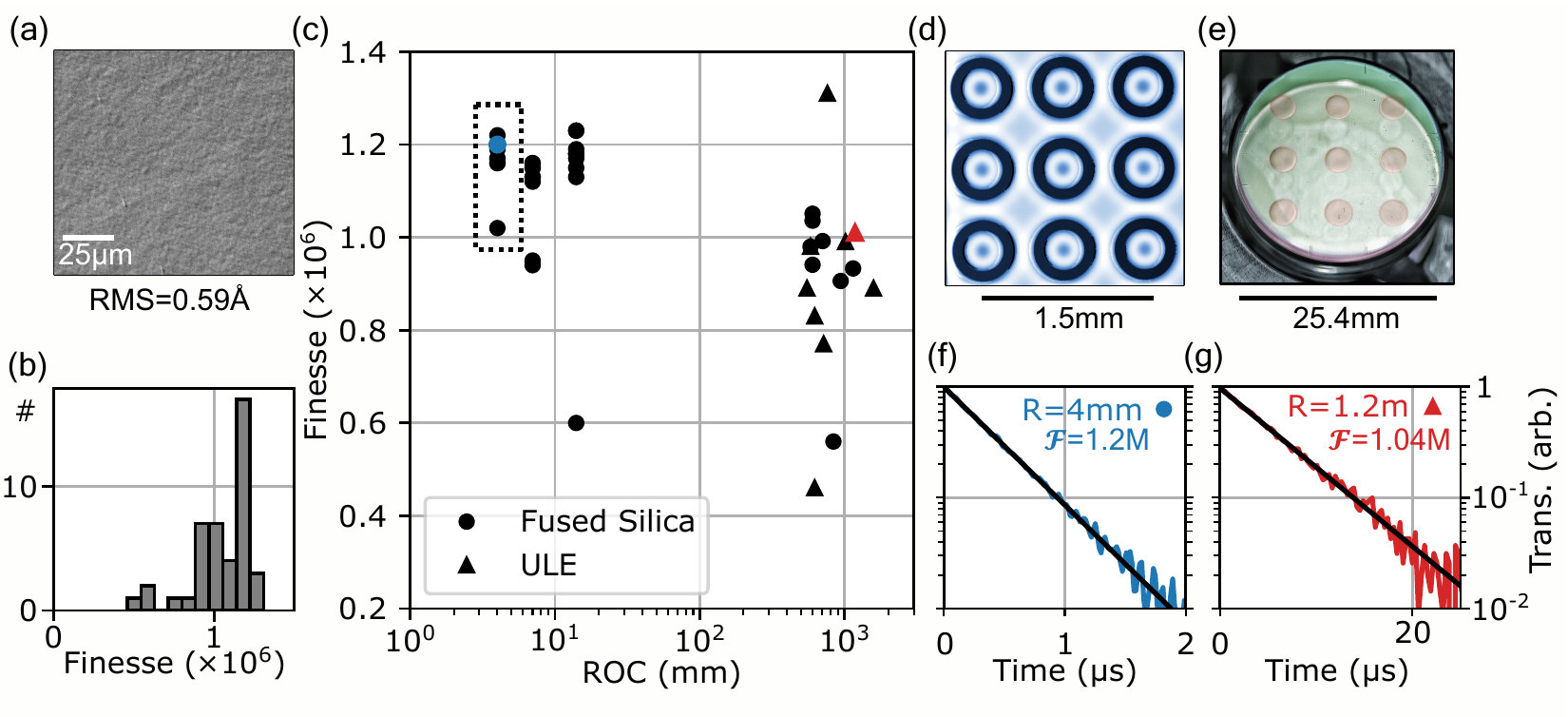}
\caption{{\bf Mirror Characterization and Cavity Performance Evaluation}
{\bf (a)} Dark-field image taken in the center of a large-$R$ ($\approx1$ m) microfabricated mirror, revealing an RMS surface roughness of 0.59 \AA.  
{\bf (b-c)} Histogram and summary of cavity finesse measurements for different $R$ micromirrors. 43 cavities, formed on 5 substrates (both fused silica and ULE) were measured.  Exemplary small- and large-$R$ cavities are highlighted in blue and red respectively, with underlying ringdown data shown in {\bf (f)} and {\bf (g)}. The dashed box corresponds to the mirror array shown in {\bf (d)}. The 9 ULE mirrors are from the substrate shown in {\bf (e)}.
{\bf (d)} Profilometry image of small-$R$ mirror array, corresponding to the dashed box in {\bf (c)}.
{\bf (e)} Image of large-$R$ mirror array on ULE substrate (triangles in  {\bf (c)}). Mirrors highlighted in false color.
{\bf (f-g)} Averaged transmission ringdown of a small-$R$ ($R=4$~mm, $L=320$~$\mu$m) and a large-$R$ ($R=1.2$ m, $L=5.5$ mm) cavity, where the light is cut off at $t=0$.  Black line is exponential fit yielding 410 ns and 6.1 $\mu$s decay time, corresponding to a finesse of 1.20 million and 1.04 million, respectively.} 
\label{f:F3}
\end{center}
\end{figure*}

The performance of these devices was evaluated by applying a state-of-the-art, ultralow-loss dielectric mirror coating, with alternating SiO$_2$/Ta$_2$O$_5$ layers designed to produce reflectivity $>0.99999$.  We then paired these substrates with flat mirrors from the same coating run, forming arrays of plano-concave Fabry-Pérot resonators.  These cavities were held in kinematic mounts, or clamped/bonded to an annular spacer.  Both small- and large-$R$ mirrors were tested, spanning mode waists from 23 $\mu$m to 220 $\mu$m. To evaluate the finesse of each resonator, a laser was mode-matched to the fundamental cavity mode, and switched off rapidly after being brought on resonance. 
Sample transmission ringdowns of small- and large-$R$ cavities are shown in Fig. \ref{f:F3}f and \ref{f:F3}g.
Exponential fits of these measurements reveal cavity lifetimes ($\tau$) of 410 ns and 6.1 $\mu$s for the small and large $R$ resonator devices, corresponding to finesse values of 1.20 million and 1.04 million (using $\F=\pi \tau c/L$, where $c$ is the speed of light and $L$ is the cavity length).
These lifetime measurements were corroborated using microwave-calibrated frequency sweep measurements.  
Note that the smallest fabricated devices (with $R\approx 100$~$\mu$m) did not receive mirror coatings, but simulations based on their surface profiles predict comparable finesse to the measured devices.
Further measurement details are available in the Supplement Section S3B.
Ringdown measurements were performed on 43 cavities created using 5 different patterned micromirror substrates, with results summarized in Fig. \ref{f:F3}b and \ref{f:F3}c.
These measurements indicate consistent performance across the fabricated samples.
The small $R$ cavities all come from a single substrate, containing a grid of 81 micromirrors.  
Out of 27 mirrors tested, 24 were found to have a finesse $>1$~million (1.13$\pm$0.13 million).
The large $R$ cavities show slightly increased variability (0.91$\pm$0.20 million), but still reach a maximum finesse of 1.31 million.  
This variability is likely due to the increased tilt sensitivity of large-mode-waist cavities, which places more stringent requirements on the mirror symmetry and cavity alignment.
We also note that the large $R$ devices are fabricated on both fused silica and ultra-low-expansion (ULE) glass, confirming compatibility with these two technologically important materials\cite{matei_15um_2017, fortier_generation_2011}.

While the finesse permits us to quantify the total mirror loss (\tran+\absb+\scat), it is also instructive to separate the different loss contributions.  
Since both mirrors of all tested microcavities were simultaneously coated, receiving an identical multilayer coating, it is reasonable to assume that the transmission coefficients (\tran) are identical for both mirrors. 
With this assumption, we can extract (\absb+\scat) from the relative transmitted and reflected powers on resonance\cite{hood_characterization_2001}.  Doing so, we estimate $\tran=1.9$ ppm for this coating, which means that, for our measured $\F =1.2\times10^6$, we infer the excess loss to be $(\absb+\scat)\approx0.74$ ppm.  Note that since these dissipative loss channels are smaller than the external loss (\tran), this resonator technology offers a path to efficient light extraction at these ultrahigh finesse levels.
\begin{figure*}[hhtt]
\begin{center}
\includegraphics[width=.4\textwidth]{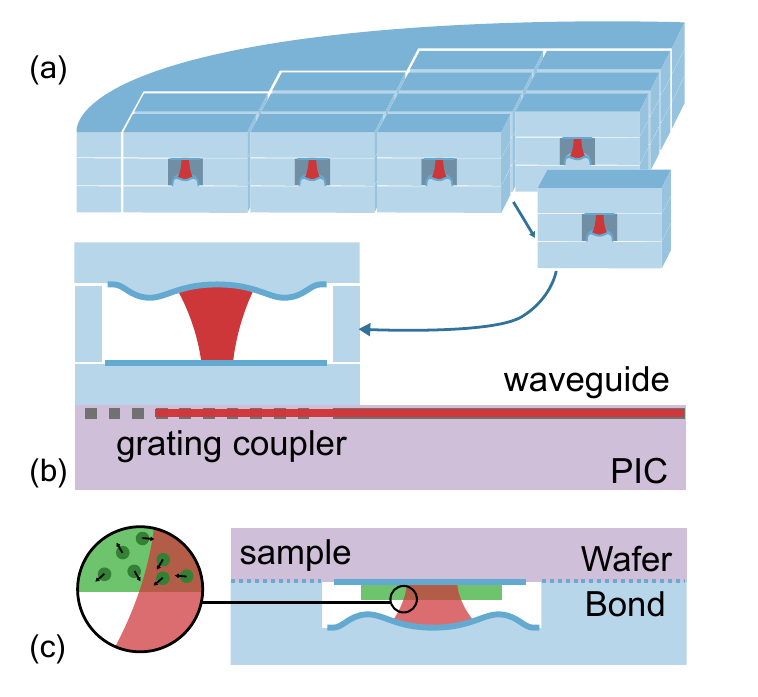}
\caption{{\bf Integrated Microcavity Outlook}
{\bf (a)} Illustration of large-scale micro-Fabry-Pérot assembly and integration, in which a planar mirror wafer is bonded to a micromachined spacer layer and a micromirror array wafer. {\bf (b-c)} Illustration of possible applications: {\bf (b)} Integration of micro-Fabry-Pérot with photonic integrated circuit (PIC). {\bf c} Bonding of recessed mirrors 
to form low-volume resonators for cavity QED.
\label{f:F4}}
\end{center}
\end{figure*}

Building on these techniques, one could envision using wafer-scale fabrication approaches (pictured in Fig. 4) to bring the unique advantages offered by high-finesse Fabry-Pérot resonators to integrated systems. In contrast to dielectric waveguide resonators, the modes of Fabry-Pérot resonators can be engineered to live almost entirely in vaccum, avoiding problematic sources of thermorefractive noise produced by dielectrics\cite{gorodetsky2004fundamental,braginsky2000thermo}. For this reason, ultrahigh-finesse cavities, of the type fabricated in this paper, could prove instrumental to satisfy the growing demand for frequency stabilized ultra-narrow linewidth lasers for atomic clocks, communications, and sensing applications. For such applications, large mode sizes ($\gtrsim 200$~$\mu$m) produced by larger radius of curvature ($\gtrsim 1$~m) mirrors are used to suppress residual noise generated by the mirror coating. Such frequency-stabilized cavities are typically constructed from ultra-low-expansion (ULE) glass to eliminate expansion-induced frequency drift. Through a separate study, a bonded cavity assembly, using 1 m-ROC micromirrors from Fig. \ref{f:F3}c, was shown to produce a thermal-noise-limited fractional frequency instability of $7\times10^{-15}$ at 1 second, in a volume of only 8 mL\cite{mclemore_thermal}. This same device was also used to lock an integrated semiconductor laser using the Pound-Drever-Hall (PDH) technique, yielding $\sim 1$ Hz integral linewidth on the timescale of a second\cite{guo_towards}.

Conversely, the small-ROC mirrors can yield small mode-volumes necessary to produce enhanced coupling rates with atoms, ions, and defect centers for quantum applications\cite{janitz_fabry-perot_2015,merkel_coherent_2020,takahashi_strong_2020,albrecht_coupling_2013} (see Fig. \ref{f:F4}c).  For example, the smallest microcavities studied here ($R= 4$mm, $L = 320\mu$m) produce modes with a waist radius of 23~$\mu$m and finesse of 1.2 million, corresponding to a Purcell enhancement factor of $\sim1000$ \cite{purcell_spontaneous_1995}, which could already enable strong coupling to quantum emitters. Since the modes of such Gaussian beam resonators are readily mode-matched to optical fibers, they permit highly efficient collection of photons required for cavity QED and quantum networking applications.
 
To harness these and other performance advantages, one could envision integrating such high finesse resonators with planar photonic circuits using vertical-emission grating couplers\cite{cheng2020grating}, as seen in Fig~4b. 
In the context of integrated photonics, these high finesse resonators are also remarkable for their ability to produce very high Q-factors (Q $>$ 10 billion) within compact footprints ($\sim1$~mm$^2$)
Hence, these resonators could offer compelling performance advantages relative to state-of-the-art ring resonators\cite{puckett2021422,lee_spiral_2013} and dielectric resonators\cite{savchenkov_optical_2007}, opening the door to scalable integrated photonic technologies.

\section*{Contributions}
{P.K. developed early reflow techniques.  N.J. carried out device fabrication, with assistance from Y.L. and D.M.. Y.L. developed numerical simulation tools.  N.J., C.A.M., D.M., J.P.H., and Y.L. contributed to cavity characterization. All authors contributed to manuscript preparation. F.Q., S.A.D., and P.R. supervised and led the collaboration.}

\section*{Acknowledgements}
Device fabrication was carried out with support from the Yale SEAS Cleanroom and Yale West Campus Cleanroom. We thank Yong Sun, Sean Rinehart, Kelly Woods, Min Li, Lei Wang, Yubo Wang and Hong X. Tang for their assistance in device fabrication. We thank Lindsay Sonderhouse and Jules Stuart for helpful comments on the manuscript, and Ramin Lalezari of FiveNine Optics for assistance in characterizing mirror surface quality.  This work was supported by NIST, the DARPA A-Phi under Award FA9453-19-C0029, and U.S. Department of Energy, Office of Basic Energy Sciences, Division of Materials Sciences and Engineering under Award DE-SC0019406. This work is a contribution of an agency of the U.S. government and not subject to copyright in the USA. Identification of commercial vendors is for scientific clarity only and does not represent endorsement by NIST. 

\section*{Methods}
\textbf{Mirror fabrication} Shipley S18 series photoresist is patterned on a super-polished glass substrate provided by Coastline Optics. After priming the substrate with Hexamethyldisilazane (HMDS)\cite{li_fabrication_2004}, we reflow the photoresist by mounting the substrate in the top of a home-made chamber filled with propylene glycol methyl ether acetate (PGMEA) vapor. The vapor is heated to approximately $45^{\circ}$C and the substrate is kept at an elevated temperature of approximately $50^{\circ}$C, so that the photoresist gradually undergoes reflow without being dissolved by the vapor. When the resist disks reach the desired shape, we stop the reflow by removing the substrate from the chamber and baking out the excess solvent. This shape is then transferred into the substrate with SF$_6$/Ar-based reactive ion etching. High reflectivity coatings are applied by FiveNine Optics Inc. Further details are available in Supplementary Information.

\textbf{Mirror characterization} 
Micromirror profiles are characterized using a Zygo Nexview.
These profiles are used as input for simulation tools based on numerical beam propagation and an eigenmode solver to estimate scattering loss ($\scat$ in main text).
In finesse measurements, cavity arrays are formed by pairing micromirrors with flat mirrors coated simultaneously. 
Their optical lifetimes are determined through ring-down measurements, where the decay of transmitted light is recorded after switching off a resonant excitation laser.
Cavity free-spectral ranges are either measured by scanning a tunable laser or inferred from cavity length.
In addition to finesse, following Ref\cite{hood_characterization_2001}, we are able to determine the excess loss (\scat+\absb) of each mirror by measuring the resonant transmission and reflection. 

\let\oldaddcontentsline\addcontentsline
\renewcommand{\addcontentsline}[3]{}
\let\addcontentsline\oldaddcontentsline

\newpage
\renewcommand{\figurename}{{\bf Fig.}}
\renewcommand{\tablename}{{\bf Table}}
\setcounter{figure}{0}\renewcommand{\thefigure}{{\bf S\arabic{figure}}}
\setcounter{table}{0}\renewcommand{\thetable}{{\bf S\arabic{table}}}
\setcounter{equation}{0}\renewcommand{\theequation}{S\arabic{equation}}

\section*{\LARGE Supplementary Information}
\tableofcontents

\begin{sidewaystable}
\section{Literature Survey}
\begin{center}
\begin{tabular}{|c|c|c|c|c|c|c|c|} 
 \hline
 Publication & Approach & Platform & $R$ & Finesse & Roughness & Main Text Ref\\ 
 \hline\hline
 Wachter 2019\cite{wachter_silicon_2019} & Chemical etch & Silicon wafer  & 100 - 300 $\mu$m & 500k @ 1550 nm & 2\AA & 19 \\ 
 \hline
 Fait 2021\cite{fait_high_2021} & Chemical etch & Silicon wafer & 70, 100 $\mu$m & 500k @ 1280 nm & & 21\\
 \hline
 Biedermann 2010\cite{biedermann_ultrasmooth_2010} & Chemical etch & Silicon wafer  & ~250$\mu$m & 64k @ 780 nm & 2.2\AA & 20\\
 \hline
 Hunger 2010\cite{hunger_fiber_2010} & Laser ablation & Silica fiber & \makecell{Measured: 100, 350 $\mu$m \\ Projected: 40 $\mu$m - 2 mm} & \makecell{Measured: 37k @ 780 nm \\ Projected: 130k @ 780 nm} & 2\AA & 10\\
 \hline 
 Colombe 2007\cite{colombe_strong_2007} & Laser ablation & Silica fiber & 150, 450 $\mu$m & 37k @ 780 nm& & 4\\
 \hline
 Hunger 2012\cite{hunger_laser_2012} & Laser ablation & Silica fiber &  20 $\mu$m - 2 mm & \makecell{Only mirror templates \\ Projected: 100k @ 830 nm} & & 15\\
 \hline
 Muller 2010\cite{muller_ultrahigh-finesse_2010} & Laser ablation & Silica fiber & 30 $\mu$m & 150k @ 920 nm & Below 4\AA & 14\\
 \hline
 Uphoff 2015\cite{uphoff_frequency_2015} & Laser ablation & Silica fiber &  120 - 600 $\mu$m & 190k @ 780 nm & & 16\\
 \hline
 Takahashi 2014\cite{takahashi_novel_2014} & Laser ablation & Silica fiber & 100 - 700 $\mu$m & 40k-60k @ 866 nm & & 17\\
 \hline
 Takahashi 2020\cite{takahashi_strong_2020} & Laser ablation & Silica fiber & ~560 $\mu$m & 50k @ 866 nm & & 23\\
 \hline
 Rempe 1992\cite{rempe_measurement_1992} & Traditional polishing & Glass substrate &  173 mm & 1.9M @ 850 nm & & 12\\
 \hline
 
\end{tabular}
\end{center}
\end{sidewaystable}

\newpage

\newpage
\section{Mirror Fabrication}
\label{sec:SIMirrorFab}
Leveraging hydrodynamics, we reflow lithographically defined photoresist disks into desired mirror shapes with atomic-scale surface roughness. Using carefully engineered reactive ion etching (RIE), we transfer those shapes into substrates while maintaining its smoothness. Here, we discuss details on how we achieve mirror radii of curvature ($R$) spanning from 100 $\mu$m to 1 m while maintaining surface roughness at sub-Angstrom level.

\subsection{Reflow Apparatus}
\label{sec:SI_ReflowApparatus}

\begin{figure}[hhtt]
\centering
\includegraphics[width=0.9\textwidth]{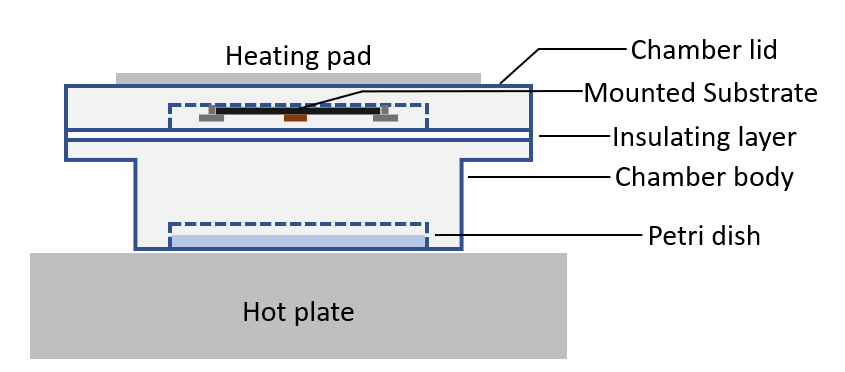}
\caption{\label{fig:Fig1}\textbf{Reflow apparatus schematic}}
\label{fig:SI-Fig1}
\end{figure}

Because the desired shape of the reflowed photoresist is an \textit{intermediate} state, precise control of the reflow is required to consistently obtain mirror surfaces with the desired $R$. To meet this requirement, we built a dedicated chamber to perform reflow under well-controlled vapor pressure and temperature. Fig. \ref{fig:SI-Fig1} shows a schematic of this apparatus. A petri dish filled with PGMEA is placed in the chamber to generate vapor, and a substrate with patterned photoresist is mounted on the chamber lid. 
An insulating layer between the body and lid allows us to maintain a temperature difference between the two parts.
By using a hot plate underneath the chamber and a heating pad attached to the lid, we are able to independently control the temperature of the solvent vapor (and thus its vapor pressure) and the substrate. This control ensures a gradual reflow while preventing the solvent from dissolving the photoresist. 
Typical temperature settings for the solvent vapor and substrate would be approximately $45^{\circ}$C and $50^{\circ}$C, respectively.  Under this condition, a 3 mm-diameter photoresist disk will reach a concave shape in approximately one hour.
As long as we maintain adequate thermalization of this apparatus, we are able to consistently reproduce the same resist shape given identical initial disk diameters and reflow times. 

\subsection{Reflow strategy: Large $R$}
\label{sec:SI_LargeR}

\begin{figure}[hhtt]
\centering
\includegraphics[width=\textwidth]{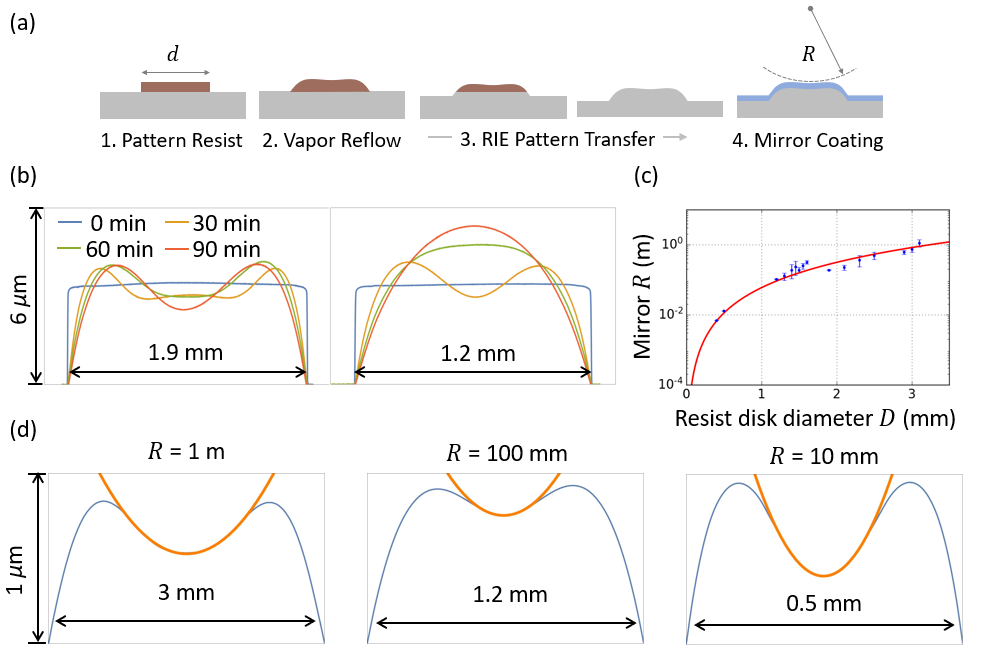}
\caption{\label{fig:Fig2-ReflowEvo}\textbf{Reflow of single-level photoresist patterns} (a) Fabrication sequence for single-level photoresist disks. (b) Shape evolution of single-level photoresist disks under reflow. Two disks with different diameters were reflowed under the same conditions and their shapes were checked at the at 30 minute intervals. At each point, the reflow was paused and excess solvent was baked out. The left panel shows a 1.9 mm diameter disk which reflows into the desired concave shape with a large aperture at 60 min. The right panel shows a 1.2~mm diameter disk which reflows completely into a convex shape after 90 minutes. (c) An empirical relation between the  mirror radius of curvature $R$ and the diameter of resist disks $D$. (d) Linecuts of mirror profiles produced by reflowed single-level photoresist.}
\label{fig:SI-Fig2}
\end{figure}

For large-$R$ mirrors, we begin with a single-level photoresist disk (Fig. \ref{fig:SI-Fig2}(a)).
Fig. \ref{fig:SI-Fig2}(b) shows measured resist cross sections at different intermediary times during reflow.  Initially, the disk redistributes its central volume outward, rounding the disk edge and forming two humps ($t=30$ min in the left panel of Fig. \ref{fig:SI-Fig2}(b)). As reflow proceeds, the humps gradually move inward, eventually merging and forming a concave surface in the middle ($t=60$ min in the left panel of Fig. \ref{fig:SI-Fig2}(b)). The humps continue merging, eventually filling in the concave valley and finally reaching a convex shape (as shown in the right panel of Fig. \ref{fig:SI-Fig2}(b)).

As shown in Fig. \ref{fig:SI-Fig2}(b), for a given disk diameter, the point where two humps just start to merge yields the largest aperture. This is the point at which we typically interrupt the reflow by taking the substrate out of the chamber and baking out the excess solvent. After obtaining the desired resist shape, we transfer it into the substrate by reactive ion etching (RIE). Details of our optimized RIE recipe will be presented below, but for geometric control purposes, we note that nonidentical etch rates of the substrate and photoresist will result in a vertical rescaling of the photoresist pattern. 
If we define the etch rate ratio of resist to substrate as $s$ ($\approx 4$ for our etch recipe), then we find the curvature scales accordingly: $R_{\mathrm{mirror}}=sR_{\mathrm{resist}}$

Under the above protocol, by sweeping the diameter of photoresist disks $D$ from 400 $\mu$m to 3~mm, we are able to achieve $R$ from 5 mm to 1 m. Fig. \ref{fig:SI-Fig2}(d) shows a gallery of example mirrors with various $R$ we have fabricated. Fig. \ref{fig:SI-Fig2}(c) summarizes an empirical relation between the mirror $R$ and resist diameter $D$ (resist height 2.5 $\mu$m), where the red line is a power law fitting yielding $R[\rm m]=0.059D[\rm mm]^{2.4}$. The error in $R$ given a particular $D$ comes from slightly different thermal conditions and reflow timing between samples. The exact relation can in principle be calculated from hydrodynamics. We note that in principle, even small-diameter disks can reach arbitrarily large $R$ as they transitions from concave to convex, but this is an unreliable way to reach large-$R$, and results in insufficient mirror apertures.

The above fabrication process, using single-level photoresist patterns, can already offer a wide range of $R$ values and in principle reach $R$ even smaller than 5 mm. For applications targeting ultra-small mode volumes, further shrinking the diameter of single-level resist disks may still be a viable approach to achieving small mirror $R$. 
For instance, a single-level, 60 $\mu$m-diameter resist disk can yield a mirror with $R=$ 750 $\mu$m, but the optical mode size must remain below 6 $\mu$m to achieve ultrahigh finesse at $\lambda=$1550 nm.
Effectively, this maximum spot size $w$, along with $R$, sets an upper bound for the cavity length $L$. If we want to exceed this cavity length limit, it is necessary to have more control over the resist shape, as discussed further below.

\subsection{Reflow Strategy: Small $R$}
\label{sec:SI_SmallR}

\begin{figure}[hhtt]
\centering
\includegraphics[width=\textwidth]{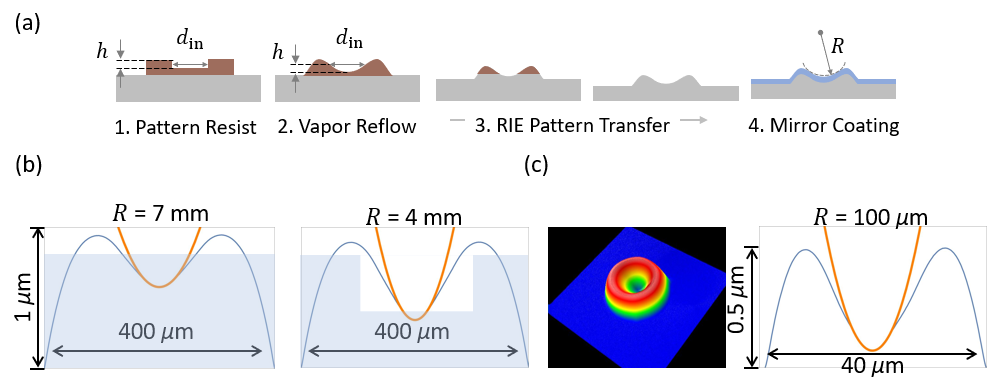}
\caption{\label{fig:Fig3}\textbf{Reflow of two-level photoresist patterns} (a) Fabrication sequence for two-level photoresist disks. (b) Linecuts of mirror profiles produced by a single-level pattern (on the left) and a two-level pattern (on the right) with the same disk diameter. (c) 3D profile and cross-section of a mirror template with the smallest $R$ fabricated in this work.}
\label{fig:SI-Fig3}
\end{figure}

In order to extend our technique towards smaller-$R$ mirrors, we modify the photoresist patterning described above. Knowing that the target shape of the reflowed resist has a central recess, we pattern a hole with diameter $d_\mathrm{in}$ and depth $h$ in the center of the initial resist disk as a zeroth-order approximation of the final shape, as shown in Fig. \ref{fig:SI-Fig3}(a). For a brief reflow, we can assume that the sharp corners of the hole simply soften, maintaining the overall dimensions and forming a concave surface in the center. Hence the aperture size and recess depth of the reflowed resist will stay around $d_\mathrm{in}$ and $h$, respectively. If we further approximate the center part of the dimple as a paraboloid, we can derive its radius of curvature to be $R_{\mathrm{resist}}=d_{\mathrm{in}}^2/8h$. Taking the etch selectivity into account, $R$ of the etched mirror will be $R=sd_\mathrm{in}^2/8h$. 
Fig. \ref{fig:SI-Fig3}(b) shows cross sections of two fabricated mirrors made from single- and two-level resist patterns with the same outer diameter. The two-level pattern results in a deeper mirror recess and smaller $R$. Fig. \ref{fig:SI-Fig3}(c) shows a 3D profile and a cross section of a mirror template with the smallest $R$ we have achieved using this two-level pattern technique ($R=100\,\mu$m).

\subsection{Aperture Constraints}
\label{sec:SI_Aperture}
\begin{figure}[hhtt]
\centering
\includegraphics[width=\textwidth]{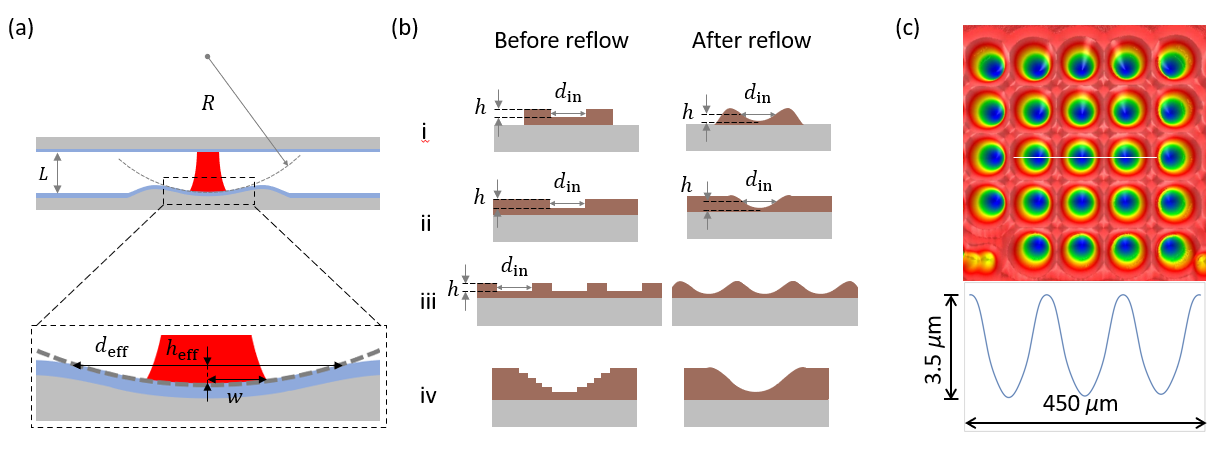}
\caption{\label{fig:Fig4} \textbf{Towards full control of mirror shapes} (a) Schematic of a plano-concave Fabry-P\'erot cavity built with our fabricated mirror. The cavity has a length of $L$, and a curved mirror with radius of curvature $R$, effective mirror aperture diameter $d_{\mathrm{eff}}$ and mirror depth $h_{\mathrm{eff}}$. The Gaussian beam radius at the curved mirror is denoted as $w$. (b) Several resist pattern strategies for better shape control and the anticipated resist shapes after reflow: (i) Two-level pattern, (ii) hole pattern, (iii) array of holes and (iv) grey-scale pattern. (c) 3D profile and linecut of a reflowed hole array following (biii).}
\label{fig:SI-Fig4}
\end{figure}

When targeting a certain $R$, the size of usable mirror aperture should go into design consideration to avoid clipping losses. This results in an important requirement on the mirror depth. 
We consider a plano-concave cavity configuration, as illustrated in Fig. \ref{fig:SI-Fig4}(a). For such a cavity to have stable modes, the cavity length $L$ must be shorter than the $R$ of the curved mirror. Under this condition, the Gaussian beam radius at the curved mirror can be expressed as 
\begin{equation}
\label{eq:design1}
    w=\left({\frac{\lambda L}{\pi}}\right)^{\frac{1}{2}}\left(\frac{L}{R}(1-\frac{L}{R})\right)^{-\frac{1}{4}}
\end{equation}
where $\lambda$ is the wavelength of light. We can estimate the clipping loss of a finite-sized curved mirror by assuming the energy in the Gaussian tail outside the aperture is totally lost, which yields 
$\mathcal{L}_\mathrm{clip}=$~$\exp\{-2(\frac{d_\mathrm{\mathrm{eff}}}{2w})^2\}$
, where $d_{\mathrm{eff}}$ is the effective aperture diameter of the mirror, as defined in Fig.~\ref{fig:SI-Fig4}(a). To maintain $\mathcal{L}_\mathrm{clip}\lesssim$ 1 ppm, we require
\begin{equation}
\label{eq:design2}
    d_\mathrm{eff}\gtrsim5w 
\end{equation}
For our nearly parabolic mirror profiles, this diameter can be linked to the mirror recess depth, $h_\mathrm{eff}$, as illustrated in Fig. 4(a):
\begin{equation}
\label{eq:design3}
    d_\mathrm{eff}=2\sqrt{2h_\mathrm{eff}R}
\end{equation}
Thus, we can translate our minimum aperture requirement (Eq. \ref{eq:design2}) into a condition linking the mirror depth and the achievable waist, or equivalently the maximum length for a given $R$ by
\begin{equation}
\label{eq:design4}
     \frac{L}{R}<\frac{A}{1+A} \quad \mathrm{with} \quad A=\left(\frac{8\pi}{25}\frac{h_\mathrm{eff}}{\lambda}\right)^2
\end{equation}
Note in this expression, the bound is totally determined by $h_{\mathrm{eff}}/\lambda$, where $\lambda$ is the wavelength of light, and it is relaxed when $h_{\mathrm{eff}}$ goes up. For reference, we typically have $h_{\mathrm{eff}}=0.3$ $\mu$m for reflowed single-level resist disks, which allows $L/R<3.7\%$ at $\lambda=1.55$ $\mu$m, and increasing the depth to $h_{\mathrm{eff}}=0.5$ $\mu$m with the two-level patterning technique enables $L/R<9.5\%$. With thicker resist films or tweaks on etch selectivity,  $h_\mathrm{eff}\simeq\lambda=1.55\,\mu$m should be within reach, which covers the design space up to the confocal configuration, i.e. $L/R\lesssim50\%$.

From the above analysis, we see that in the cavity design phase, when we are given a set of cavity parameters, (e.g. $L$ and $R$ or target mode parameters), it is wise to first calculate $h_{\mathrm{eff}}$ to minimize the clipping loss. After settling the geometric parameters $R$ and $h_{\mathrm{eff}}$, or equivalently $d_{\mathrm{eff}}$ and $h_{\mathrm{eff}}$ through Eq. \ref{eq:design3}, for a certain cavity mirror, we can implement the strategy presented in Sec. \ref{sec:SI_SmallR} with $d_{\mathrm{in}}\simeq d_{\mathrm{eff}}$ and $h\simeq s\cdot h_{\mathrm{eff}}$ to design the initial resist shape, as illustrated in Fig. \ref{fig:SI-Fig4}(bi). With this strategy, we note that the outer diameter of the two-level resist disk will influence the final shape of the recess. To reduce this complication, we can use another patterning strategy where we only define holes in a uniform layer of resist, as shown in Fig. \ref{fig:SI-Fig4}(bii). This approach also offers flexibility in making arrays of such mirrors, as illustrated in Fig. \ref{fig:SI-Fig4}(biii).  An example of such a reflowed pattern is shown in Fig. \ref{fig:SI-Fig4}(c). Finally, if we want to gain further control of mirror shapes, we can move to a grey-scale patterning technique, enabled by direct laser lithography, as illustrated in Fig. \ref{fig:SI-Fig4}(biv).

\subsection{Surface quality control}
\label{sec:SI_SurfaceQuality}
Our micro-fabricated mirrors begin with super-polished flat substrates with sub-Angstrom surface roughness, and it is crucial that our fabrication does not significantly worsen this surface figure. 
The reflow process inherently generates a smoothed mirror template, eliminating unwanted high-frequency surface texture. Therefore, we mainly need to ensure that the surface quality is not degraded by the reactive ion etch or by unwanted defects in the substrates and photoresist.

To achieve smooth transfer of the reflowed pattern, we develop an RIE recipe that preserves the surface quality of both the photoresist and substrates. Reactive ion etching of quartz/fused silica has been studied extensively. In particular, it has been shown in \cite{Li-smoothetch,minnick_optimum_2013} that a plasma based on SF$_6$ with a greater concentration of heavy noble gases (Ar or Xe) under low pressure can preserve ultrasmooth surfaces on quartz. At the same time, to avoid detrimental modification of the photoresist during etching (e.g. reticulation\renewcommand*{\citenumfont}[1]{S#1}\cite{hsieh1993reticulation}\renewcommand*{\citenumfont}[1]{#1}), we use low RF power and apply thermal grease to transfer heat between the substrate and carrier wafer. We carry out the etch using an Oxford 100 PlasmaLab, with pressure 4.5 mTorr, RF power 90 W, SF$_6$ flow rate 4 sccm, Ar flow rate 14 sccm and helium backing on. This recipe typically gives a DC offset of ~360 V with a 0.5 mm silica carrier wafer, S1818 resist etch rate of $\sim$100 nm/min and fused silica etch rate of $\sim$25 nm/min.

During the etch, defects on and inside the substrates could be exposed and further magnified by the plasma, which typically result in pits of few to tens of nanometer depth on the mirror surface. Such defects could come from surface and subsurface damage introduced by polishing processes and ion impurities in manufacture processes, and a simple way to test their presence is to etch a witness substrate without patterning. Photoresist quality is also an important factor, and we find careful handling and storage conditions to be important in avoiding defects.

Finally, we note that our fabrication flow can in principle be applied to other common materials, for example silicon \cite{kharel_ultra-high-q_2018}, GaAs and sapphire, as long as they can be polished to the necessary level and their etch recipe can be tuned to preserve surface quality.

\section{Mirror Characterization}
\subsection{Cavity mode simulation}
\label{sec:SI_Simulation}
When targeting ultrahigh finesse, it is important to consider any possible sources of loss caused by the fabrication technique.  In particular, scattering from surface defects and clipping losses due to mirror shape errors are of concern.  To guide our development process and lend confidence in the mirror template quality prior to optical coating, we employ numerical simulation techniques to predict mirror losses.  Our tools build on several previously established techniques.  First, one can utilize Fourier optics to simulate a beam undergoing repeated rounds of propagation, and monitor the rate at which energy decays (due to propagation outside the finite mirror).  Alternatively, one can employ an eigenmode analysis technique, in which a round trip of cavity propagation is encoded in a scattering matrix, whose complex eigenvalues can be calculated to find mode frequencies and loss rates.  In a separate work \renewcommand*{\citenumfont}[1]{S#1}\cite{luo_simulation}\renewcommand*{\citenumfont}[1]{#1}, we cross-check these techniques against each other and verify their ability to simulate these ultrahigh finesse values.  By inputting measured mirror profilometry data to these tools, we can simulate the expected scattering/clipping loss of the mirror profiles after etching.  In general, we find that our fabrication technique is consistently able to produce scattering/clipping losses at the ppm level.  Even for distinctly non-parabolic surfaces, we typically find that the mirrors can still support ultrahigh finesse, albeit with modes that deviate from traditional Gaussian modes.  In fact, these simulation tools can also help guide the development of alternative mirrors for intentionally forming non-Gaussian resonators.
\subsection{Finesse measurement}
\label{sec:SI_measurement}
\begin{figure}[hhtt]
\centering
\includegraphics[width=\textwidth]{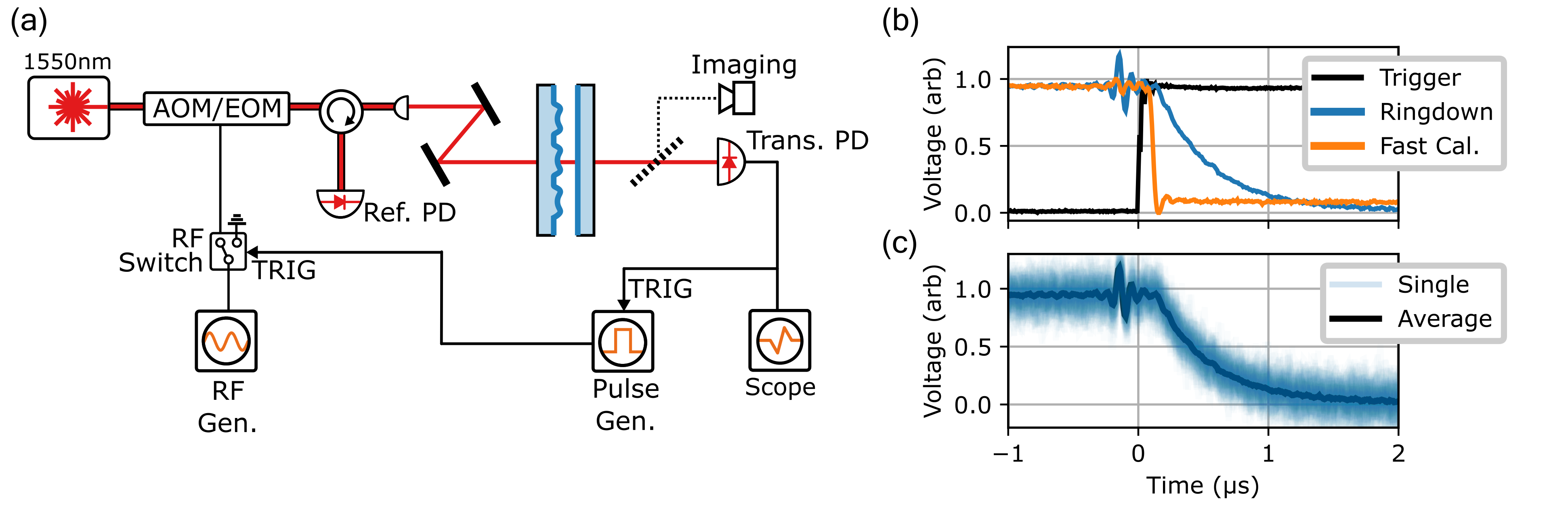}
\caption{\textbf{Ringdown Measurements} (a) Optical setup. (b) Demonstration of system response time, using a faster calibration cavity.  The ringdown of this faster cavity is seen in orange, demonstrating that the system (modulator, detector, amplifiers, triggers, etc) has sufficienty fast response time to measure our fastest ringdowns (e.g. the blue data, corresponding to one of the shortest measured cavities). (c) Example of ringdown trace averaging.
}
\label{fig:SI-Fig5}
\end{figure}

Finesse values were obtained from optical ringdowns, in which a laser is brought on resonance, then switched off abruptly.  The decay of the transmitted light can be fit to an exponential to extract the cavity lifetime.  Depending on the length of the cavity, the ringdown times varied from ~400 ns to 10s of $\mu$s.  For longer (shorter) ringdowns, the laser was switched off by switching off the RF drive to an acousto-optic modulator (electro-optic modulator), thus cutting off the laser carrier (sideband).  The RF switch was triggered by a pulse, generated when the transmission reached a certain threshold voltage, indicating that the laser was on resonance.  For the faster decay times, it is important to confirm that the response time of the whole system (the pulse generator, RF switch, optical modulator, photodetector, and any amplifiers) is sufficiently fast to measure the intended signal without distortion.  To test this, we used a separate optical resonator with a much faster decay ($\approx$5 ns), as illustrated in the calibration curve of Fig. \ref{fig:SI-Fig5}(b).  We find that the response time is $\leq$50 ns, indicating a sufficiently fast bandwidth for our measurements.  Note that, for some measurements, we also implement averaging of many ringdowns to improve SNR, as illustrated in Fig. \ref{fig:SI-Fig5}(c).

\renewcommand*{\bibnumfmt}[1]{[S#1]}

\end{document}